\documentclass[aps,prl,twocolumn,showpacs,superscriptaddress,nofootinbib]{revtex4}

\usepackage{amsfonts,epsfig}
\usepackage{latexsym}
\usepackage{epsfig}
\usepackage{amsmath}
\usepackage{amssymb}
\usepackage{amsfonts}
\usepackage{amsthm}
\usepackage{mathrsfs}
\DeclareMathAlphabet{\mathrsfs}{U}{rsfs}{m}{n}
\DeclareMathAlphabet{\mathpzc}{OT1}{pzc}{m}{it}
\DeclareMathAlphabet{\matheus}{U}{eus}{m}{n}
\DeclareMathAlphabet{\mathbbold}{U}{bbold}{m}{n}

\newcommand{\half}{\mbox{$\textstyle \frac{1}{2}$}}
\newcommand{\ket}[1]{\left | #1 \right \rangle}
\newcommand{\bra}[1]{\left \langle #1   \right |}

\newcommand{\Tr}{\operatorname{Tr}}

\newcommand{\one}{\mathbbold{1}}

\newcommand{\comment}[1]{}

\newcommand{\ba}{\begin{align}}
\newcommand{\ea}{\end{align}}

\begin{document}

\title{Security Proof for Quantum Key Distribution Using Qudit Systems}

\author{Lana Sheridan}
\affiliation{Centre for Quantum Technologies, National University of Singapore, Singapore}
\author{Valerio Scarani}
\affiliation{Centre for Quantum Technologies, National University of Singapore, Singapore}
\affiliation{Department of Physics, National University of Singapore, Singapore}

\date{March 29, 2010}


\begin{abstract}
We provide security bounds against coherent attacks for two families of quantum key distribution protocols that use $d$-dimensional quantum systems. In the asymptotic regime, both the secret key rate for fixed noise and the robustness to noise increase with $d$. The finite-key corrections are found to be almost insensitive to $d\lesssim 20$. 
\end{abstract}


\maketitle


\textit{Introduction. --} The field of quantum key distribution (QKD) comprises topics ranging from applied mathematics to technological developments \cite{review1,revdusek,review2,revlo}. In such a large field, it is normal that progress may not be homogeneous. Here we deal with a topic that was studied in detail a few years ago, then left aside, and is now coming back to the forefront: QKD protocols using systems of dimension larger than two (\textit{qudits}).

There is an obvious advantage in using high-dimensional alphabets for QKD: each signal carries $\log d>1$ bits, so a larger amount of information can be sent for a given transmission of the channel. Moreover, the first studies indicated that the resistance to noise of the protocols increases when the dimension is increased, both for one-way \cite{BT00,CBKG02} and two-way post-processing \cite{BCEEKM03,AGS03,NA05}. At the level of implementation, qudit encoding in photonic states has been demonstrated using angular momentum modes \cite{angmom} or time-bins \cite{rob}. However, at some point the interest of the community shifted towards different challenges, perceived as more urgent. As a consequence, both full security proofs and proper implementations of higher-dimensional protocols are still lacking.

In this paper, we start filling the first gap. For a wide class of higher-dimensional protocols, we provide a security bound against coherent attacks that takes into account finite-key effects. In the asymptotic limit, our bound vindicates the previous partial results concerning the higher resistance to noise. Moreover,we show that finite-key effects vary little with $d$. In this work, we assume that the signal is really a qudit; as such, our bounds cannot be immediately applied to implementations: issues like a more accurate description of the optical signal \cite{GLLP} and the squashing property at detection \cite{BML08} need to be addressed in future research.

\textit{The protocols. --} We focus on two families of protocols, both introduced first in \cite{CBKG02}: \textit{two-basis} protocols, the natural generalization of the Bennett-Brassard 1984 protocol (BB84) for qubits \cite{bb84}; and \textit{$(d+1)$-basis} protocols, the generalization of the six-state protocol for qubits \cite{sixstate1,sixstate2}.

A few reminders and notations first. The Weyl operators, a generalization of the Pauli matrices for larger dimensions, are defined by $U_{jk}=\sum_{s=0}^{d-1} \omega^{sk} \ket{s+j}\bra{s}$ for $j,k\in\{0,1,...,d-1\}$ and $\omega$ is the $d^{\text{th}}$ root of unity. The generalized Bell basis states are $\ket{\Phi_{jk}} = \sum_{s=0}^{d-1} \omega^{sk} \ket{s \ s+j}=\one\otimes U_{jk}\ket{\Phi_{00}}$. The state $\ket{\Phi_{00}}=\frac{1}{\sqrt{d}}\sum_{s}\ket{ss}$ is invariant under $U\otimes U^*$, where the star denotes complex conjugation in the computational basis.

The entanglement-based version of the protocols under study is as follows. Alice prepares $\ket{\Phi_{00}}$ and sends one of the qudits to Bob. At measurement, Alice measures in the eigenbasis of one of the $U_{jk}$ chosen at random; Bob does similarly using one of the $U^{*}_{jk}$. In the sifting phase, they keep only the items for which they used the same bases. The parameters that are estimated are the error vectors
\begin{equation}
\underline{q}_{\,jk}=\{q_{jk}^{(0)},q_{jk}^{(1)},...,q_{jk}^{(d-1)}\}
\end{equation}
where $q_{jk}^{(t)}=\mathrm{Prob}(a-b=t\mod d|j,k)$ is the probability that Alice's outcome $a$ and Bob's outcome $b$ differ by $t$, modulo $d$, when the basis of $U_{jk}$ was chosen by both. The probability of no error $q_{jk}^{(0)}= 1-\sum_{t=1}^{d-1}q_{jk}^{(t)}$ appears in the vector for convenience. Even if we do not consider this here, note that one can sometimes obtain better estimates by checking the statistics of measurements in different bases as well \cite{LKEKO03,WMU08}.

Now, there are $d^2-1$ non-trivial $U_{jk}$, but some of the corresponding eigenbases carry redundant information. The most elegant choice consists in choosing a subset of these which are \textit{mutually unbiased bases} (MUB). There are at least two and at most $(d+1)$ such bases, which explains the choice of the two protocols. Specifically, for the two-basis protocol, we can choose $U_{10}$ and $U_{01}$. 
However, a subset of the $U_{jk}$ only form a complete MUB set when $d$ is prime.  Our study of $(d+1)$-basis protocols will be restricted to these dimensions, the choice of bases being the set $\{U_{01},U_{1k} : k\in[0,d-1]\}$.

\textit{Security bounds: preliminary considerations. --} We focus on security bounds for \textit{one-way post-processing without pre-processing}. The information-theoretical formula for the secret key rate achievable against coherent attacks is known and the same for all protocols; but the most general coherent attacks are defined by an infinite number of parameters, so the formula cannot be computed directly. For most protocols, one rather relies on the following fact (see \cite{review2} for an explanation and the exceptions): the bound for coherent attacks is asymptotically the same as the one for \textit{collective attacks}, which are defined by a small number of parameters.

The two bounds, for coherent and collective attacks, are usually identical only asymptotically. The application of the same reduction to finite-key bounds requires an estimate of the difference. The exponential De Finetti theorem \cite{rennerthesis} provides such an estimate, which is however far from tight and leads to exceedingly pessimistic bounds. Among qubit protocols, much tighter estimates have been obtained for the BB84 and the six-state protocol, based on their high symmetries \cite{KGR05,RGK05}. The obvious extension of the same argument applies for the protocols under study here. Indeed, first, the parameters $\underline{q}_{\,jk}$ do not change if, before the measurement, $U_{j'k'}$ is applied on Alice's qudit and simultaneously $U_{j'k'}^{*}$ is applied on Bob's qudit. This observation follows from $[U_{jk}\otimes U_{jk}^*,U_{j'k'}\otimes U_{j'k'}^*]=0$, a consequence of $U_{jk} U_{j'k'} = \omega^{kj'-jk'} U_{j'k'} U_{jk}$. Second, the generalized Bell states are the eigenstates of all the $U_{jk}\otimes U_{jk}^*$. From there, one follows the same reasoning as in \cite{KGR05,RGK05}. So, it follows from this construction that $\rho_{AB}$ is diagonal in the generalized Bell basis:
\begin{equation}
\rho_{AB} = \sum_{j,k=0}^{d-1} \lambda_{jk} \ket{\Phi_{jk}}\bra{\Phi_{jk}}\label{belldiag}
\end{equation}
where $\sum_{j,k=0}^{d-1} \lambda_{jk} =1$. For such a state, the link with the error vector is given by
\begin{eqnarray}
q_{01}^{(t)}=\sum_{k=0}^{d-1} \lambda_{t,k} &,& q_{1k}^{(t)}=\sum_{j=0}^{d-1} \lambda_{j, (k j - t)\mod d}\,,   \label{eq:qs2}
\end{eqnarray}
which are always valid at least for $k=0$ and valid for all $k$ when $d$ is prime. Equivalently,
\begin{equation}
\lambda_{jk} = \frac{1}{d} \left(\sum_{s} q_{1s}^{(sj-k \mod d)} + q_{01}^{(j)} - 1\right)\,.
\label{eq:lambda}
\end{equation} 

\textit{Asymptotic bounds. --} For asymptotic bounds, one can assume without loss of generality that only one basis is used for the key and is chosen almost always, while the other bases are chosen with negligible probability and used to bound the eavesdropper's information \cite{LCA}. With this argument, one removes the overhead due to the sifting factor $\frac{1}{d}$ that would be present in a symmetric protocol. Here we choose the key-basis to be the one of $U_{01}$.

Eve's information is quantified by the Holevo bound $\chi(A:E|\rho_{AB}) = S(\rho_E)-\sum_{a=0}^{d-1}p(a) S(\rho_{E|a})$ where the $a$'s are the outcomes of Alice's measurement in the key-basis and where Eve is supposed to hold a purification of $\rho_{AB}$. In particular, for the Bell-diagonal state (\ref{belldiag}) one has $p(a)=\Tr(\rho_{A}\Pi_{01}^{(a)})=\frac{1}{d}$ and $S(\rho_E) = H(\underline{\lambda})$. In order to compute the $S(\rho_{E|a})$, one starts from a purification of $\rho_{AB}$: $\ket{\psi}_{ABE}=\sum_{j,k} \sqrt{\lambda_{jk}} \ket{\Phi_{jk}}_{AB}\ket{e_{jk}}_E$ where $\ket{e_{jk}}_E$ is an arbitrary orthonormal basis for Eve's system. Bob's system is traced out, then Alice makes projections onto her part of the remaining system in the computational basis, leading to $\rho_{E|a} = \Tr(\rho_{AE}\Pi_{01}^{(a)})/p(a)$. These matrices are found to have a block-diagonal structure with different eigenvectors but same eigenvalues, leading to $S(\rho_{E|a})=H(\underline{q}_{01})$ for all $a$. In summary,
\begin{equation}
\chi(A:E|\underline{\lambda}) = H(\underline{\lambda}) - H(\underline{q}_{01}) \,.
\end{equation}
For the $(d+1)$-basis protocols with $d$ prime, the $\underline{\lambda}$ are uniquely determined by the $\underline{q}_{jk}$ through Eq.~(\ref{eq:lambda}), so Eve's information is $I_E=\chi(A:E|\underline{\lambda})$. For the 2-basis protocols, Eve's information must be taken as $I_E=\max\chi(A:E|\underline{\lambda})$ where the maximum is taken over all choices of $\underline{\lambda}$ compatible with the observed error vectors $\underline{q}_{01}$ and $\underline{q}_{10}$.

To do this, we parameterize the $\lambda$s:
\begin{equation}
\lambda_{j,(d-k)} = a_j^{(k)} q_{01}^{(j)} \,,
\end{equation}
where $\sum_{k} a_{j}^{(k)} = 1 \ \forall j$.
From equation~(\ref{eq:qs2}), $q_{10}^{(t)} = \sum_{j=0}^{d-1} \lambda_{j,(d- t)}$.  So, we have the set of constraints $q_{10}^{(t)} = \sum_{j=0}^{d-1} a_j^{(t)} q_{01}^{(j)}$.  To minimize $I_E$, for each $t$ all $a_j^{(t)}$ must be equal and equal to $q_{10}^{(t)}$.  Then since $H(\underline{\lambda}) = H(\underline{q}_{01}) + \sum_t q_{01}^{(t)} H(\underline{a}_t)$ and $\underline{a}_t = \underline{q}_{10} \ \forall t$ we have
\begin{equation}
I_E = H(\underline{q}_{10}) \,.
\end{equation}

As a concrete \textit{a priori} benchmark, we assume that the observation yields the natural generalization of the qubit depolarizing channel:
\begin{equation}
\underline{q}_{\,jk}\,\equiv\,\underline{q}_{\,jk}{(Q)}=\{1-Q,Q/(d-1),...,Q/(d-1)\}\label{qq} 
\end{equation}
for all bases $j,k$ observed in the protocol. In the case of $(d+1)$-basis protocols, this fixes $\lambda_{00}=1-\frac{d+1}{d}Q$ and all the others $\lambda_{jk}=Q/d(d-1)$, leading finally to
\begin{align}
\hspace{-0.8em} I_E(Q) &= -(1 - \frac{d+1}{d} Q) \hspace{-0.3em} \left(\log (1 - Q - \frac{Q}{d}) - \log (1 - Q)\right) \nonumber \\
 & \hspace{-0.8em} - \frac{Q}{d} \left(\log \frac{Q}{d^2-d} - \log (1-Q)\right) - Q \log \frac{1}{d}\,.\label{iedbases}
\end{align} 
In the case of 2-basis protocols,
\begin{equation}
I_E(Q)=-Q \log \frac{Q}{d-1} - (1 - Q) \log (1 - Q)\,\equiv\,H(\underline{Q})\,. \label{ie2bases}
\end{equation}
Note that the corresponding $\rho_{AB}$ can be obtained from $\ket{\Phi_{00}}$ by passing Bob's qudit through the optimal asymmetric universal, resp. phase covariant, $1\rightarrow 2$ cloner \cite{CBKG02}. The secret key fraction is given by $r_{\infty}= \log d - H(\underline{Q}) - I_E(Q)$. The critical values of $Q$ at which $r_{\infty}$ becomes zero are given in Table~\ref{tab:rford}.

\begin{table}
  \begin{tabular}{| c || c | c | }
       \hline
    $\ \ d \ \ $ & $ \qquad Q_{\text{2-basis}} \qquad$ & $\quad Q_{\text{$(d+1)$-basis}} \quad$  \\ \hline
    2 & 11.00 & 12.62 \\
    3 & 15.95 & 19.14 \\ 
    4 & 18.93 & 23.17  \\ 
    5 & 20.99 & 25.94  \\
    7 & 23.72 & 29.53  \\
    11 & 26.82 & 33.36  \\
    \hline
  \end{tabular}
   \caption{Value of $Q$ at which $r_{\infty}=0$ for 2-basis and $(d+1)$-basis protocols, assuming one-way post-processing without pre-processing.  \label{tab:rford}
}
\end{table}

The result (\ref{ie2bases}) was already presented as Eq.~(22) in~\cite{CBKG02} as a lower bound. It was obtained by means of an entropic uncertainty relation developed by Hall~\cite{Hall}. Strictly speaking, this relation involves the classical mutual information and as such cannot be used for security against collective attacks. However, the same relation was recently shown to hold for Holevo quantities \cite{RB09,BCCRR09}: so the bound derived using entropic uncertainty relations is ultimately correct, and is tight for the 2-basis protocols.

\textit{Finite key bounds. --} We consider now the realistic case where $N<\infty$ signals have been exchanged, following \cite{SR08,CS09}. In this case, all the steps of the protocols have some probability of failure. For error correction and privacy amplification, these probabilities are denoted by $\varepsilon_{EC}$ and $\varepsilon_{PA}$ respectively; the estimate of any measured parameter $V$ may fail with probability $\varepsilon_{PE}$ and the law of large numbers implies that one has to consider a fluctuation $\Delta V = \Delta V(\varepsilon_{PE})$. In addition to those, as mentioned above, the mathematical estimates using smooth Renyi entropies may fail with probability $\bar{\varepsilon}$. The security parameter is the total probability of failure 
$
\varepsilon = \varepsilon_{EC}+\varepsilon_{PA} + n_{PE}\varepsilon_{PE} +\bar{\varepsilon}  ,
$
where $n_{PE}$ is the number of parameters estimated in the protocol (for simplicity, we assume the same error on all parameters).

With all these notions in place, the lower bound for the secret key rate reads\footnote{In the final term of this expression, the factor $(2 \log d+3)$ appears.  Starting from~\cite{SR08} and propagating to other finite keys papers, this was mistakenly given as $(2d+3)$, which for qubits is 7, rather than 5 as it should be.  Therefore this is an inconsequential change for qubits, but for higher dimensions however, the difference to the bound can be more significant.  The origin of this term is explained in~\cite{rennerthesis}.}
\begin{eqnarray}
r_{N} &=&  \frac{n}{N} \left(\min_{E|\mathbf{V\pm\Delta V}} H(A|E) - H(A|B)-\frac{1}{n} \log\frac{2}{\varepsilon_{EC}}\right.   \nonumber  \\
& & \    \left. -\frac{2}{n} \log \frac{1}{\varepsilon_{PA}}- (2 \log d+3) \sqrt{\frac{\log(2/\bar{\varepsilon})}{n}}\right) .
\label{eq:rn}
\end{eqnarray} 
The origin of each term should be clear from the failure probabilities and has been discussed in detail in previous work \cite{SR08,CS09}; we have not put any overhead on the efficiency of error correction. The term $n/N$ describes the fact that only $n<N$ signals can be devoted to create a key, because some signals must be used for parameter estimation. We have $\min_{E|\mathbf{V\pm\Delta V}} H(A|E)=\log(d) - I_E$; $I_E$ is given by (\ref{iedbases}) or (\ref{ie2bases}), in which the ``true" values ${q}_{\,jk|\infty}^{(t)}$ are estimated by the worst case values ${q}_{\,jk|m}^{(t)}={q}_{\,jk|\infty}^{(t)}\pm\Delta q_{jk}^{(t)}$ compatible with the fluctuations. Obviously, the worst case is defined by increasing the errors ($t\in\{1,...,d-1\}$) and decreasing ${q}_{\,jk|m}^{(0)}$ correspondingly in order to preserve the normalization of probabilities.

Now, for each given value of $N$, $\varepsilon$ and $\varepsilon_{EC}$, one has to maximize $r_N$ by the best choice of the other parameters of the protocol: here, the probabilities $p_{jk}$ of choosing each basis (supposed the same for Alice and Bob) and the failure probabilities. This is done numerically. For simplicity, we keep using only the basis $U_{01}$ for the key, so $n=Np_{01}^2$ (we have checked that the improvement obtained by taking all the bases is rather negligible).

A subtle difference with the qubit case appears in the treatment of statistical fluctuations. Consider the basis $j,k$ and suppose that $m=Np_{jk}^2$ signals have been measured in this basis by both Alice and Bob: the law of large numbers provides the bound
\begin{align}
||\underline{q}_{\,jk|m}-&\underline{q}_{\,jk|\infty}|| = \sum_{t=0}^{d-1}|\Delta q_{jk|m}^{(t)}| \leq \xi(m,d) \,,\nonumber \\
 \text{where } \ & \xi(m,d)=\sqrt{\frac{2\ln(1/\varepsilon_{PE})+2 d \ln(m+1)}{m}}\,.
\end{align} 
The only additional constraint is the normalization $\sum_{t=0}^{d-1}\Delta q_{jk|m}^{(t)}=0$. So, if $d>2$, we cannot find a tight bound for each $\Delta q_{jk|m}^{(t)}$, $t\in\{1,...,d-1\}$. In one extreme case, only one $q_{jk|m}^{(t')}$ carries all the fluctuations, leading to $\Delta q_{jk|m}^{(t)}=\half\xi(m,d)\,\delta_{t,t'}$; in the other extreme case, all the fluctuations of the error values are identical i.e. $\Delta q_{jk|m}^{(t)}=\frac{1}{2(d-1)}\xi(m,d)$. It turns out that this last case provides slightly most conservative bounds, so the graphs are plotted for this case; we also checked that the brute bound $\Delta q_{jk|m}^{(t)}=\half\xi(m,d)$ for all $t$ is definitely too pessimistic.

\begin{figure}[h!]
\begin{center}
\includegraphics[width=3in]{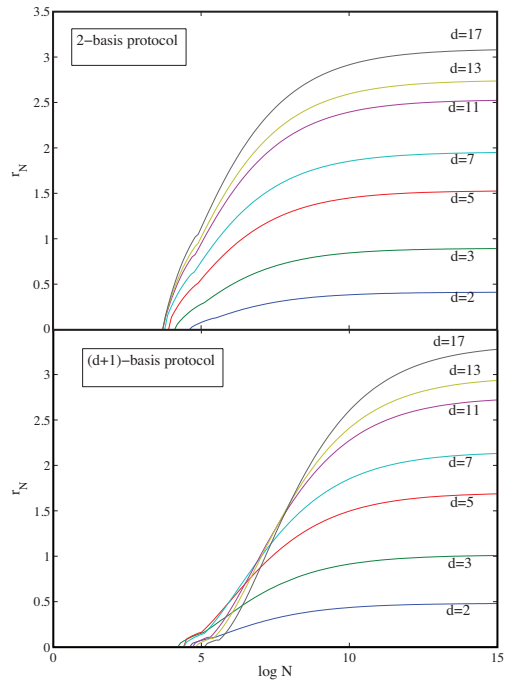}
\caption{Secret key rate as a function of the number of signals $N$ for $\varepsilon=10^{-5}$, $\varepsilon_{EC}=10^{-10}$ and $Q=5\%$. Above: 2-basis protocols; Below: $(d+1)$-basis protocols.}
\label{fig:finiterates}
\end{center}
\end{figure}

Having addressed these concerns, we are now able to run the numerical optimizations. Since we are providing \textit{a priori} estimates, we assume the observed error vectors to be $\underline{q}_{\,jk}{(Q)}$ given in (\ref{qq}). Also, for the $(d+1)$-basis case, we fixed $p_{1k}=\frac{1-p_{01}}{d}$. The results are shown in Figure \ref{fig:finiterates}. The dominant finite-key correction is the one due to the statistical fluctuations, which goes as $\xi(m,d)\sim\sqrt{d}$ rather than linearly in $d$: this explains why, for the dimensions we plotted, the critical value is always around $N\sim 10^{5}$. 

\textit{Conclusion. --} We have provided security bounds against coherent attacks for QKD protocols that use higher-dimensional alphabets, that are valid in the non-asymptotic regime of finite-length keys. When choosing either the secret key rate or the robustness to noise as the figure of merit, this study confirms that higher-dimensional protocols perform better than the corresponding qubit protocols.


\begin{acknowledgments}
This work was supported by the National Research Foundation and the Ministry of Education, Singapore.

\end{acknowledgments}


\bibliography{QKDbib8}

\end{document}